\documentclass[11pt,a4paper]{article}

\usepackage[utf8]{inputenc}
\usepackage[T1]{fontenc}
\usepackage{lmodern}
\usepackage{amsmath,amssymb,mathtools}
\usepackage{geometry}
\usepackage{microtype}
\usepackage{bm}
\usepackage{cite}
\usepackage{hyperref}

\newtheorem{prop}{Proposition}[section]
\newtheorem{remark}{Remark}

\hypersetup{colorlinks=true,linkcolor=blue,citecolor=blue,urlcolor=blue}

\title{\textbf{Equilibrium Thermodynamic Properties of Two-Component and Two-Phase Mixtures}}
\author{S.~Benjelloun}
\date{Devinci Higher Education, DeVinci Research Center, Paris, France}

\begin{document}
\maketitle

\begin{abstract}
We develop a general thermodynamic framework for determining the equilibrium thermodynamic and thermophysical properties of systems composed of two constituents or two phases. The formulation does not require the introduction of a specific equation of state for the constituent phases or for the mixture and can therefore be applied using either analytical equations of state, tabulated thermodynamic data, or experimental measurements of the constituent properties. Two classes of systems are considered. The first consists of two immiscible constituents in complete thermodynamic equilibrium, with no mass transfer between them. The second consists of two phases of the same substance in equilibrium, with mass exchange between the phases, as occurs in liquid--vapor systems and blackoil models.

For both classes of systems, we derive expressions for the mixture compressibility, equilibrium speed of sound, thermal expansion coefficients, heat capacities, and Gr\"uneisen coefficient in terms of the corresponding properties of the individual constituents or phases. Particular attention is given to two-phase equilibrium, for which pressure and temperature are constrained by the coexistence relation and cannot be treated as independent variables. The resulting expressions provide a consistent alternative to simple arithmetic averaging of thermophysical properties and can be applied directly to experimentally measured or tabulated phase properties. The framework is illustrated for ideal-gas mixtures and for liquid--vapor equilibrium, including asymptotic limits relevant to cavitating flows.
\end{abstract}


\section{Introduction}
Thermodynamic properties of mixtures and multiphase systems are required in a wide range of applications involving compressible flows, phase transitions, cavitation, acoustics, and energy conversion. In practical calculations, the properties of a mixture are frequently obtained from an equation of state constructed specifically for the mixture, or, in simplified approaches, by averaging the properties of the individual constituents. Although such procedures can be useful in particular applications, they may introduce unnecessary assumptions and may obscure the thermodynamic constraints imposed by equilibrium. These approaches may also be inconvenient when the properties of the individual constituents are available directly from experiments, tabulated data, or independent equations of state.

A particularly important application is provided by black-oil systems,
in which liquid and gas phases coexist under thermodynamic equilibrium and
mass may be transferred from one phase to the other. Water-vapor systems are also a common example of equilibrium two-phase thermodynamics with mass exchange. Such systems illustrate the essential difficulty of equilibrium two-phase thermodynamics: the pressure and temperature of the phases are not independent variables along
the coexistence curve, while the thermodynamic properties of the mixture depend on the phase fraction.  Our work also addresses systems without mass exchange, such as air--water mixtures and other immiscible
two-phase mixtures.

A thermodynamic system at equilibrium can be described by a suitable set of independent state variables. The remaining state functions and their derivatives are related through thermodynamic identities \cite{Benjelloun2021}. Consequently, once a sufficiently complete set of thermodynamic properties is known, the other thermodynamic coefficients can be obtained without necessarily constructing an explicit equation of state for the mixture.

The purpose of this work is to develop such a formulation for systems consisting of two constituents or two phases. The approach is based directly on the equilibrium thermodynamic identities and on the additive nature of the extensive specific quantities of the mixture. It therefore does not depend on a particular mixture equation of state. This feature is especially useful when the constituent properties are available from experimental measurements or tabulated thermodynamic databases. Two physically distinct cases are considered. First, we consider two immiscible constituents that are in mechanical and thermal equilibrium but do not exchange mass. Second, we consider two phases of the same substance in complete thermodynamic equilibrium, with mass transfer between the phases. A liquid--vapor system provides the canonical example. While expressions for the equilibrium speed of sound in particular two-phase systems have been reported previously, the present formulation places these results within a common thermodynamic framework and derives, in addition, the associated heat capacities, thermal expansion coefficients, and Gr\"uneisen coefficient. The formulation is designed to be used directly with analytical, tabulated, or experimentally measured
phase properties.

\section{Notation}
Throughout this paper, lower-case letters denote specific quantities
upper-case letters denote the corresponding thermodynamic coefficients or
quantities when appropriate. The density is denoted by $\rho=1/v$, where
$v$ is the specific volume. The notation used for the principal
thermodynamic and thermophysical quantities is summarized in
Table~\ref{tab:notation}.

\begin{table}[htbp]
\centering
\caption{Notation for the thermodynamic and thermophysical quantities used
throughout the paper.}
\label{tab:notation}
\renewcommand{\arraystretch}{1.25}
\begin{tabular}{lll}
\hline
\textbf{Symbol} & \textbf{Definition} & \textbf{Description} \\
\hline
$P$ & -- & Pressure \\
$T$ & -- & Temperature \\
$e$ & -- & Specific internal energy \\
$s$ & -- & Specific entropy \\
$h$ & $h=e+pv$ & Specific enthalpy \\
$\mu$ & -- & Chemical potential \\
\hline
$c$ &
$c^2=\left(\dfrac{\partial p}{\partial\rho}\right)_s$ &
Adiabatic (equilibrium) speed of sound \\
$c_T$ &
$c_T^2=\left(\dfrac{\partial p}{\partial\rho}\right)_T$ &
Isothermal speed of sound \\
$C_v$ &
$C_v=T\left(\dfrac{\partial s}{\partial T}\right)_v$ &
Specific heat at constant volume \\
$C_p$ &
$C_p=T\left(\dfrac{\partial s}{\partial T}\right)_p$ &
Specific heat at constant pressure \\
$\gamma$ &
$\gamma=C_p/C_v$ &
Ratio of specific heats \\
$k$ & $dp=c^2\,d\rho+\rho kT\,ds$ & Gr\"uneisen coefficient \\
$\alpha_p$ & $\displaystyle \alpha_p=\frac{1}{v} \left(\frac{\partial v}{\partial T}\right)_p$ & Isobaric thermal expansion coefficient \\
$\chi_T$ & $\displaystyle \chi_T=-\frac{1}{v} \left(\frac{\partial v}{\partial p}\right)_T = \frac{1}{\rho c_T^2}$ & Isothermal compressibility \\
$\alpha_v$ & $\alpha_v= \left(\frac{\partial P}{\partial T}\right)_{\rho} = \rho k C_v$ & Isovolume caloric coefficient \\
\hline
$L$ & $L=h_G-h_L=T(s_G-s_L)$ & Latent heat of phase change \\
\hline
\end{tabular}
\end{table}

For a two-constituent mixture, subscripts $1$ and $2$ refer to the two
constituents, whereas for a two-phase system subscripts $G$ and $L$ denote
the gas and liquid phases, respectively. 


The principal thermodynamic coefficients are related by the identities \cite{Benjelloun2021}
\begin{align}
\chi_T &= \frac{1}{\rho c_T^2},\\
C_v &= C_p-\frac{T\alpha_p^2}{\rho c_T^2},\\
c^2 &= c_T^2\frac{C_p}{C_v} = \gamma \, c_T^2,\\
k &= \frac{\alpha_p c_T^2}{C_v} = \frac{\alpha_p c^2}{C_p} ,\\
\alpha_v &= \rho k C_v
          =\rho\alpha_p c_T^2.
\end{align}



\section{Main Results}

We distinguish between mixtures without mass exchange and mixtures undergoing phase change. The two cases lead to qualitatively different equilibrium constraints.

\subsection{Immiscible equilibrium mixture}

Consider a closed mixture of two immiscible constituents, denoted by $1$ and $2$, at mechanical and thermal equilibrium, with common pressure $P$ and temperature $T$, and with constant mass fractions
\[
x_1=x,\qquad x_2=1-x.
\]
Let $v_i,\rho_i,$ denote the specific volume, and density of constituent $i$, and $v, rho$ the specific volume, and density of the mixture with
$$v=xv_1+(1-x)v_2,$$
$$\frac{1}{\rho}=\frac{x}{\rho_1}+\frac{1-x}{\rho_2}.$$

\begin{prop}
Let $c_{T,i},C_{p,i},\alpha_{p,i}$ denote the isothermal speed of sound, specific heat at constant pressure, and
thermal expansion coefficient of constituent $i$.

Then the mixture properties are given by
\begin{align}
C_p&=xC_{p,1}+(1-x)C_{p,2},
\end{align}
and
\begin{equation}
\alpha_p=
\rho\left(
\frac{x\alpha_{p,1}}{\rho_1}
+\frac{(1-x)\alpha_{p,2}}{\rho_2}
\right).
\end{equation}

\begin{equation}
\frac{1}{\rho^2c_T^2}
=
\frac{x}{\rho_1^2c_{T,1}^2}
+\frac{1-x}{\rho_2^2c_{T,2}^2}
\end{equation}

The remaining equilibrium coefficients follow from the identities \cite{Benjelloun2021}
\begin{align}
C_v&=C_p-\frac{T\alpha_p^2}{\rho c_T^2},\\
c^2&=c_T^2\frac{C_p}{C_v},\\
k&=\frac{\alpha_p c_T^2}{C_p}\\
\alpha_v &= \rho k C_v   =\rho\alpha_p c_T^2.
\end{align}

In particular, the equilibrium adiabatic sound speed is
\begin{equation}
\boxed{
\frac{1}{\rho^2c^2}
=
\frac{x}{\rho_1^2c_{T,1}^2}
+\frac{1-x}{\rho_2^2c_{T,2}^2}
-
\frac{
T\left(
\dfrac{x\alpha_{p,1}}{\rho_1}
+\dfrac{(1-x)\alpha_{p,2}}{\rho_2}
\right)^2
}{
xC_{p,1}+(1-x)C_{p,2}
}.
}
\end{equation}

\end{prop}

\begin{remark}
Using the relation between isothermal and adiabatic compressibility, the equilibrium speed of sound can be written in terms of the adiabatic sound speeds:
\begin{equation}
\boxed{
\begin{aligned}
\frac{1}{\rho^2c^2}
={}&
\frac{x}{\rho_1^2c_1^2}
+\frac{1-x}{\rho_2^2c_2^2}
-\frac{xT\alpha_{p,1}^2}{\rho_1^2C_{p,1}}
-\frac{(1-x)T\alpha_{p,2}^2}{\rho_2^2C_{p,2}}\\
&-
\frac{
T\left(
\dfrac{x\alpha_{p,1}}{\rho_1}
+\dfrac{(1-x)\alpha_{p,2}}{\rho_2}
\right)^2
}{
xC_{p,1}+(1-x)C_{p,2}
}.
\end{aligned}
}
\end{equation}
\end{remark}

\begin{remark}
We see in particular that the speed of sound $c^2$ does not depend simply on $c_1$ and $c_2$ but on the other thermophysical properties as well. The formulae are equivalent to formulae given in \cite{landau1989course, temkin1992sound}. Here we have developed formulae for all the other thermodynamic quantities for the mixture, \textit{i.e.}, $\alpha_v$, $\alpha_p$, $k$, $C_v$, $C_p$, $\gamma$, etc.
\end{remark}

\subsection{Equilibrium two-phase mixture with mass exchange}

Consider a two-phase system composed of a gas phase $G$ and a liquid phase
$L$, in complete thermodynamic equilibrium and with mass exchange between
the phases. Let
\[
x=\frac{m_G}{m_G+m_L}
\]
be the gas mass fraction. Along the saturation curve, pressure and
temperature are related by the Clapeyron equation
\begin{equation}
\frac{dP}{dT}
=
\frac{L}{T(v_G-v_L)}
=
\frac{s_G-s_L}{v_G-v_L},
\end{equation}
where
\begin{equation}
L=h_G-h_L=T(s_G-s_L)
\end{equation}
is the latent heat of phase change.

\begin{prop}
Define
\begin{align}
K_G={}&
\frac{\gamma_G^2v_G^2}{c_G^2}
-\alpha_Gv_G\frac{2T(v_G-v_L)}{L}
+\frac{TC_{p,G}(v_G-v_L)^2}{L^2},\\
K_L={}&
\frac{\gamma_L^2v_L^2}{c_L^2}
-\alpha_Lv_L\frac{2T(v_G-v_L)}{L}
+\frac{TC_{p,L}(v_G-v_L)^2}{L^2}.
\end{align}

Then the equilibrium sound speed is
\begin{equation}
\boxed{
c^2=
\frac{v^2}{xK_G+(1-x)K_L}.
}
\end{equation}

The corresponding Gr\"uneisen coefficient and heat capacity at constant
volume are
\begin{align}
\boxed{
k=
\frac{v}{L}\,
\frac{xK_G+(1-x)K_L}{v_G-v_L},
}
\\[1ex]
\boxed{
C_v=
\frac{L^2}
{T(v_G-v_L)^2}
\left[xK_G+(1-x)K_L\right].
}
\end{align}

On the coexistence curve, $P$ and $T$ are not independent. Consequently,
the usual coefficients involving independent variations of both $P$ and
$T$ are singular; in particular,
\[
C_p=\infty,\qquad
\alpha_p=\infty,\qquad
\gamma=\infty,
\]
while the combinations entering $c$, $k$, and $C_v$ remain finite.
\end{prop}

\section{Derivation and Proof of Proposition 3.1.}

We first consider two immiscible constituents at complete thermodynamic equilibrium. There is no mass exchange, so the mass fractions $x$ and $1-x$ remain constant. The mixture extensive quantities per unit total mass are therefore
\begin{align}
v&=xv_1+(1-x)v_2, \label{eq:v}\\
e&=xe_1+(1-x)e_2,\\
s&=xs_1+(1-x)s_2 \label{eq:s}.
\end{align}

Since both phases have the same $P$ and $T$, their first-law identities
\begin{align}
de_1&=Tds_1-Pdv_1,\\
de_2&=Tds_2-Pdv_2
\end{align}
combine directly into
\begin{equation}
de=Tds-Pdv.
\end{equation}

\subsection{Isothermal compressibility}

At constant temperature, we derive \eqref{eq:v}
\begin{equation}
-\left(\frac{\partial v}{\partial P}\right)_T
=
x\left[-\left(\frac{\partial v_1}{\partial P}\right)_T\right]
+
(1-x)\left[-\left(\frac{\partial v_2}{\partial P}\right)_T\right].
\end{equation}
Using
\begin{equation}
-\left(\frac{\partial v_i}{\partial P}\right)_T
=
\frac{1}{\rho_i^2c_{T,i}^2},
\end{equation}
we obtain
\begin{equation}
\boxed{
\frac{1}{\rho^2c_T^2}
=
\frac{x}{\rho_1^2c_{T,1}^2}
+
\frac{1-x}{\rho_2^2c_{T,2}^2}.
}
\end{equation}

\subsection{Thermal expansion and heat capacity}

Differentiation at constant pressure gives
\begin{equation}
\left(\frac{\partial v}{\partial T}\right)_P
=
x\left(\frac{\partial v_1}{\partial T}\right)_P
+
(1-x)\left(\frac{\partial v_2}{\partial T}\right)_P.
\end{equation}
Since
\begin{equation}
\left(\frac{\partial v_i}{\partial T}\right)_P
=v_i\alpha_{p,i}
=\frac{\alpha_{p,i}}{\rho_i},
\end{equation}
we obtain
\begin{equation}
\boxed{
\alpha_p=
\rho\left(
\frac{x\alpha_{p,1}}{\rho_1}
+\frac{(1-x)\alpha_{p,2}}{\rho_2}
\right).
}
\end{equation}

Similarly, from \eqref{eq:s}
\begin{equation}
\left(\frac{\partial s}{\partial T}\right)_P
=
x\frac{C_{p,1}}{T}
+(1-x)\frac{C_{p,2}}{T},
\end{equation}
and hence
\begin{equation}
\boxed{
C_p=xC_{p,1}+(1-x)C_{p,2}.
}
\end{equation}

\subsection{Sound speed and completion of the proof}

The thermodynamic identities connecting $c$, $c_T$, $C_p$, $C_v$, and
$\alpha_p$ give
\begin{equation}
C_v=C_p-\frac{T\alpha_p^2}{\rho c_T^2},
\end{equation}
and
\begin{equation}
c^2=c_T^2\frac{C_p}{C_v}.
\end{equation}
Consequently,
\begin{equation}
\frac{1}{\rho^2c^2}
=
\frac{1}{\rho^2c_T^2}
-\frac{T\alpha_p^2}{\rho^2C_p}.
\end{equation}
Substitution of the mixture expressions for $c_T$, $\alpha_p$, and $C_p$
gives exactly the expression stated in Proposition 3.2.

The result shows that the mixture sound speed contains a thermal coupling
term. Therefore, an arithmetic averaging rule for $c_1$ and $c_2$
cannot in general reproduce the correct equilibrium sound speed.





\section{Derivation and Proof of Proposition 3.2}

We now consider two phases of the same constituent that coexist and may
exchange mass. Let phase $G$ denote the gas and phase $L$ the liquid. At
equilibrium,
\begin{equation}
T_G=T_L=T,\qquad
P_G=P_L=P,\qquad
\mu_G=\mu_L.
\end{equation}
Mass conservation gives
\begin{equation}
m_G+m_L=m_{\rm tot}.
\end{equation}
Thus the system has two independent degrees of freedom. Because $P$ and
$T$ are constrained by phase equilibrium, they cannot be used simultaneously
as independent variables.

The mixture quantities are
\begin{align}
v&=xv_G+(1-x)v_L,\\
s&=xs_G+(1-x)s_L,\\
e&=xe_G+(1-x)e_L.
\end{align}

The equality of chemical potentials implies
\begin{equation}
h_G-h_L=T(s_G-s_L)=L,
\end{equation}
where $L$ is the latent heat. Differentiating the equilibrium condition gives the Clausius--Clapeyron relation
\begin{equation}
\frac{dP}{dT}
=
\frac{s_G-s_L}{v_G-v_L}
=
\frac{L}{T(v_G-v_L)}.
\end{equation}

Moreover, the equilibrium state variables $v,s,e$ verify the thermodynamic identity
$$
de = Tds - P dv.
$$ 
We can therefore define the thermodynamic coefiscients $c^2$, $k$, $C^v$, etc, and the different identities and relations apply.
 
\subsection{Equilibrium thermodynamic constraints}

Because $P=P(T)$ along saturation, derivatives involving $P$ and $T$ simultaneously as independent variables cease to be meaningful. Instead, the equilibrium coefficients must be evaluated along the saturation
manifold.

Writing,
\begin{equation}
dp=\frac{dP}{dT}\,dT,
\end{equation}
and using the thermodynamic identity
\begin{equation}
dT=\frac{T}{C_v}\,ds+\frac{kT}{\rho}\,d\rho,
\end{equation}
we obtain
\begin{equation}
dp=
\frac{dP}{dT}
\left(
\frac{T}{C_v}\,ds
+
\frac{kT}{\rho}\,d\rho
\right).
\end{equation}
On the other hand, by definition of the equilibrium speed of sound and the Gr\"uneisen coefficient,
\begin{equation}
dp=c^2\,d\rho+\rho kT\,ds.
\end{equation}
$ds$ must be identical. 
We therefore obtain
\begin{align}
c^2 &=
\frac{kT}{\rho}\frac{dP}{dT},
\label{eq:c2_dPdT}\\
\rho kT &=
\frac{T}{C_v}\frac{dP}{dT}.
\label{eq:dPdT_kCv}
\end{align}
We reformulate these two relations as 
\begin{equation}
\frac{dP}{dT}=\rho k C_v = \frac{\rho c^2}{k T},
\label{eq:dPdT_rhokCv}
\end{equation}
and
\begin{equation}
c^2=k^2TC_v.
\label{eq:c2_k2TCv}
\end{equation}


Combining these relations and applying the Clausius--Clapeyron relation gives then
\begin{equation}
 \label{rel1b}
k=
\frac{c^2v}{LT(v_G-v_L)}.
\end{equation}
and 
\begin{equation}
\label{rel2b}
C_v=
\frac{v^2L^2}
{Tc^2(v_G-v_L)^2}.
\end{equation}

\subsection{Evaluation of the equilibrium sound speed}

It remains now to find the equilibrium speed of sound $c^2$. Write
\begin{align}
v(x,P)&=xv_G(P)+(1-x)v_L(P), \label{eq:GL1}\\
s(x,P)&=xs_G(P)+(1-x)s_L(P). \label{eq:GL2}
\end{align}

Switching to the thermodynamical variables $P$ and $s$ to describe the system on the saturation curve, we can write \footnote{We write $v (x,p)=\tilde{v}(s(x,p),p)$ and we derive with respect to $x$ on both sides}
\begin{equation}
\frac{\partial v}{\partial P}\bigg\rvert_{\mathrm{sat},x}=\frac{\partial v}{\partial P}\bigg\rvert_{\mathrm{sat},s}+\frac{\partial v}{\partial s}\bigg\rvert_{\mathrm{sat},P}\frac{\partial s}{\partial P}\bigg\rvert_{\mathrm{sat},x}.
\end{equation}
Thus
\begin{equation}
\frac{\partial v}{\partial P}\bigg\rvert_{\mathrm{sat},s}=\frac{\partial v}{\partial P}\bigg\rvert_{\mathrm{sat},x}-\frac{\frac{\partial v}{\partial x}\bigg\rvert_{\mathrm{sat},P}}{\frac{\partial s}{\partial x}\bigg\rvert_{\mathrm{sat},P}}\frac{\partial s}{\partial P}\bigg\rvert_{\mathrm{sat},x}
\end{equation}
and hence using \eqref{eq:GL1}-\eqref{eq:GL2}
\begin{equation}
\frac{\partial v}{\partial P}\bigg\rvert_{\mathrm{sat},s}=\frac{\partial v}{\partial P}\bigg\rvert_{\mathrm{sat},x}-\frac{v_G-v_L}{s_G-s_L} \frac{\partial s}{\partial P}\bigg\rvert_{\mathrm{sat},x},
\end{equation}
which yields to
\begin{equation}\label{eq:dvdpsat}
\begin{split}
\frac{\partial v}{\partial P}\bigg\rvert_{\mathrm{sat},s}=&x \left(\frac{\partial v_G}{\partial P} \bigg\rvert_{\mathrm{sat},x}-\frac{v_G-v_L}{s_G-s_L} \frac{\partial s_G}{\partial P}\bigg\rvert_{\mathrm{sat},x} \right)\\&+(1-x) \left(\frac{\partial v_L}{\partial P}\bigg\rvert_{\mathrm{sat},x}-\frac{v_G-v_L}{s_G-s_L} \frac{\partial s_L}{\partial P}\bigg\rvert_{\mathrm{sat},x} \right)
\end{split}
\end{equation}
Noting $L=h_G-h_L=T(s_G-s_L)$ the latent heat of phase change.
The first term of the right hand side of Eq.~\eqref{eq:dvdpsat} can be expressed as follows
\begin{equation}\label{eq:dvdpsatrhs}
\begin{split}
\frac{\partial v_G}{\partial P}\bigg\rvert_{\mathrm{sat},x}-& \frac{v_G-v_L}{s_G-s_L}\frac{\partial s_G}{\partial P}\bigg\rvert_{\mathrm{sat},x}=
\frac{\partial v_G}{\partial P}\bigg\rvert_{\mathrm{sat},T}+\frac{\partial v_G}{\partial T}\bigg\rvert_{\mathrm{sat},P}\frac{\partial T}{\partial P}\bigg\rvert_{\mathrm{sat},x} \\
&-\frac{T (v_G-v_L)}{L} \frac{\partial s_G}{\partial P}\bigg\rvert_{\mathrm{sat},T}-\frac{T(v_G-v_L)}{L}\frac{\partial s_G}{\partial T}\bigg\rvert_{\mathrm{sat},P}\frac{\partial T}{\partial P}\bigg\rvert_{\mathrm{sat},x}
\end{split}
\end{equation}
Clapeyron's relation gives $\frac{\partial P}{\partial T}\bigg\rvert_{\mathrm{sat},x}=\frac{L}{T(v_G-v_L)}=\frac{s_G-s_L}{v_G-v_L}$, and therefore the right hand side of Eq.~\eqref{eq:dvdpsatrhs} becomes
\begin{equation}
\begin{split}
&\frac{\partial v_G}{\partial P}\bigg\rvert_{\mathrm{sat},T}+\frac{\partial v_G}{\partial T}\bigg\rvert_{\mathrm{sat},P}\frac{T (v_G-v_L)}{L}\\
&-\frac{T (v_G-v_L)}{L} \frac{\partial s_G}{\partial P}\bigg\rvert_{\mathrm{sat},T}-\frac{T^2 (v_G-v_L)^2}{L^2}\frac{\partial s_G}{\partial T}\bigg\rvert_{\mathrm{sat},P}
\end{split}
\end{equation}
Combining the fourth relation of Maxwell $\frac{\partial s}{\partial P}\bigg\rvert_{T}=-\frac{\partial v}{\partial T}\bigg\rvert_{P}$ and the definition of $C^p$ as $C^p=T\frac{\partial s}{\partial T}\bigg\rvert_{P}$, we obtain 
\begin{equation}
\frac{\partial v_G}{\partial P}\bigg\rvert_{\mathrm{sat},T}+\frac{\partial v_G}{\partial T}\bigg\rvert_{\mathrm{sat},P}\frac{2 T (v_G-v_L)}{L}-\frac{T (v_G-v_L)^2  C^p_G}{L^2}
\end{equation}
The second term of Eq.~\eqref{eq:dvdpsat} is transformed in the same way and we get
\begin{equation}
\begin{split}
-\frac{v^2}{c^2}&=x \left(\frac{\partial v_G}{\partial P}\bigg\rvert_{\mathrm{sat},P}+\frac{\partial v_G}{\partial T}\bigg\rvert_{\mathrm{sat},P}\frac{2 T(v_G-v_L)}{L}-\frac{T (v_G-v_L)^2 C^p_G}{L^2}\right)+\\
&(1-x)\left(\frac{\partial v_L}{\partial P}\bigg\rvert_{\mathrm{sat},T}+\frac{\partial v_L}{\partial T}\bigg\rvert_{\mathrm{sat},P}\frac{2 T(v_G-v_L)}{L}-\frac{T (v_G-v_L)^2C^p_L}{L^2}\right)
\end{split}
\end{equation}
and
\begin{equation}
\begin{split}
-\frac{v^2}{c^2}&=x \left(-\frac{\gamma_G ^2v_G^2}{c_G^2}+\alpha_Gv_G\frac{2T(v_G-v_L)}{L}-\frac{T (v_G-v_L)^2C^p_G}{L^2}\right)+\\
&(1-x)\left(-\frac{\gamma_L^2v_L^2}{c_L^2}+\alpha_L v_L\frac{2T(v_G-v_L)}{L}-\frac{T (v_G-v_L)^2C^p_L}{L^2}\right),
\end{split}
\end{equation}
which allows to write
\begin{equation}\label{eq:c2res}
c^2=\frac{v^2}{x\mathcal{K}_1+(1-x)\mathcal{K}_2},
\end{equation}
where
\begin{equation}
\begin{cases}
\mathcal{K}_1=\frac{\gamma_G ^2v_G^2}{c_G^2}-\alpha_Gv_G\frac{2T(v_G-v_L)}{L}+\frac{T (v_G-v_L)^2C^p_G}{L^2}\\
\mathcal{K}_2=\frac{\gamma_L^2v_L^2}{c_L^2}-\alpha_L v_L\frac{2T(v_G-v_L)}{L}+\frac{T(v_G-v_L)^2C^p_L}{L^2}
\end{cases}.
\end{equation}

This formula is coherent with the one given in \cite{landau1989course}. From the equations \eqref{eq:c2res}, \eqref{rel1b} and \eqref{rel2b} we can deduce :
\begin{equation}
k=\frac{v}{\frac{L}{v_G-v_L}\left(x\mathcal{K}_1+(1-x)\mathcal{K}_2\right)},
\end{equation}
and
\begin{equation}
C^v=\frac{L^2}{T\left(v_G-v_L\right)^2}\left(x\mathcal{K}_1+(1-x)\mathcal{K}_2\right).
\end{equation}
All the other thermodynamic coefficents can be expressed from the three above $c^2$, $C^v$ and $k$.
%




\section{Limiting Liquid--Vapor Regime}

For a liquid--vapor system with
\[
v_G\gg v_L,
\]
the general expression simplifies as follows :

\begin{equation}
c^2=\frac{\left(x+\left(1-x\right) \frac{v_1}{v_2}\right)^2}{x \left(\frac{\gamma_G ^2}{c_G^2}-\frac{2 \alpha_G  T}{L}+\frac{TC^p_G}{L^2} \right)+\mathcal{O}\left(x \frac{v_1}{v_2}\right)+(1-x)  \left(\frac{T C^p_L}{L^2}\right)+\mathcal{O}\left((1-x) \frac{v_1}{v_2}\right)}.
\end{equation}
For the case $ x\ll 1$, and still assuming that $v_G\gg v_L $ we end up with
\begin{equation}
c^2=\frac{L^2}{T C_L^p} \left(x+\frac{v_L}{v_G}\right)^2+\mathcal{O}\left(\frac{v_L^3}{v_G^3}\right)+\mathcal{O}\left(x\frac{v_L^2}{v_G^2}\right).
\end{equation}
Hence,
\begin{equation}
c\simeq
\left(x+\frac{v_L}{v_G}\right)
\frac{L}{\sqrt{TC_{p,L}}}.
\end{equation}
The corresponding coefficients are
\begin{align}
k\simeq
\left(x+\frac{v_L}{v_G}\right)
\frac{L}{TC_{p,L}},
\\
C_v\simeq C_{p,L}.
\end{align}

The equilibrium sound speed is therefore very small near the cavitation regime. This behavior is consistent with the strong reduction of measured sound speed in liquid--gas mixtures approaching cavitation \cite{Kieffer1977}. %
The identity $C^v \simeq C^p_L  $ may be surprising but one should keep in mind that the mixture coefficents, such as $C^v$, are derivatives along the saturation line. It is also not at all intuitive, and we note that in reference \cite{Kieffer1977}, the author uses the hypothesis $C^v \simeq C^v_L$ (for $x \ll 1$) which we proved above to not be generally correct.  

The opposite limit, $1-x\ll1$, leads to a gas-dominated asymptotic regime. The corresponding expressions follow by retaining the leading gas-phase
terms in the general formulas.
\begin{equation}
\begin{cases}
c^2=\frac{c_G^2}{\gamma_G ^2-c_G^2 \left(\alpha_G   \frac{2 T}{L}+\frac{T  C^p_G}{L^2}  \right)}+\mathcal{O}\left(\frac{v_L}{v_G}\right)+\mathcal{O}(1-x)
\\
k=\frac{c_G^2}{L \gamma_G ^2-c_G^2 \left(2 T \alpha_G+\frac{T  C^p_G}{L}  \right)}+\mathcal{O}\left(\frac{v_L}{v_G}\right)+\mathcal{O}(1-x) \simeq \frac{c^2}{L}
\\
C^v \simeq \frac{\gamma_G ^2  L^2}{Tc_G^2}-2 \alpha_G  L-C^p_G 
\end{cases}
\end{equation}

\section{Conclusions}

We have derived a general thermodynamic framework for equilibrium mixtures
of two constituents or phases. The formulation avoids the need to construct
a dedicated equation of state for the mixture and instead expresses the
equilibrium properties in terms of the thermodynamic properties of its
constituents.

For immiscible mixtures without mass exchange, the equilibrium volume, heat
capacity, thermal expansion coefficient, and compressibility follow from
weighted combinations of the corresponding constituent properties. The
sound speed, however, contains an additional thermal coupling term and
therefore cannot generally be obtained from a simple averaging rule.

For two-phase mixtures with mass exchange, the equality of chemical
potentials constrains pressure and temperature to the saturation curve, and
the equilibrium sound speed, Gr\"uneisen coefficient, and heat capacity must
be defined as derivatives along this curve. The resulting expressions
depend explicitly on the latent heat and the difference in specific volume
between the two phases.

The framework is applicable both to analytical equations of state and to experimentally or numerically tabulated thermophysical properties. In
particular, it provides a direct route to calculating equilibrium properties in liquid--vapor systems and applications such as black-oil and water-vapor models, as well as immiscible mixtures such as foam systems, without introducing an additional mixture equation of state.

\end{document}